%% file: main.tex
\documentclass[letterpaper]{article}
\usepackage[draft]{aaai2026}
\usepackage{times}
\usepackage{helvet}
\usepackage{courier}
\usepackage[hyphens]{url}
\usepackage{graphicx}
\usepackage{natbib}
\usepackage{caption}
\title{The Conversation Turns First:\\Crowd Discussion and Price Reversals in Prediction Markets}

\author{
    \begin{minipage}[t]{0.47\textwidth}
        \centering
        \normalfont\normalsize
        {\Large\bfseries Omneya Sultan}\\[0.3em]
        Information Sciences Institute\\
        University of Southern California\\
        Marina del Rey, California 90292, USA\\
        osultan@usc.edu
    \end{minipage}
    \hfill
    \begin{minipage}[t]{0.47\textwidth}
        \centering
        \normalfont\normalsize
        {\Large\bfseries Fred Morstatter}\\[0.3em]
        Information Sciences Institute\\
        University of Southern California\\
        Marina del Rey, California 90292, USA\\
        fredmors@isi.edu
    \end{minipage}
}
\affiliations{}

\begin{document}
\maketitle
\input{abstract}
\input{introduction}
\input{related_work}
\input{data_selection}
\input{sweep}
\input{methods}
\input{results}
\input{conclusion}
\input{acknowledgements}
\bibliography{references}
\end{document}

%% file: abstract.tex
\begin{abstract}
Prediction markets combine trading with public discussion of the same events. We examine whether comment-derived signals predict subsequent activity, buying direction, and changes in the leading outcome. A correlation sweep across 79 non-political Polymarket markets guides six classification experiments comparing comment features, trading features, and their combinations. On live blocks containing comments, attention nearly matches trading history in predicting heavy trading within 18 hours (PR-AUC 0.786 versus 0.790, against prevalence 0.606), with its relative advantage concentrated in short markets. Comment content carries directional information: toxicity ranks future buying direction above chance in 43 of 53 scored markets, while adding attention, sentiment, and stance to flow history increases ROC-AUC from 0.780 to 0.788. Leadership changes are predicted primarily by market state. Stance shifts against the leader before reversals in 23 of 27 evaluable markets, but adds no clear improvement in individual-block forecasting. These results distinguish attention from directional support and price uncertainty. They establish predictive associations consistent with discussion and trading responding to shared information, without identifying a causal effect of comments on markets.
\end{abstract}

%% file: introduction.tex
\section{Introduction}
\label{sec:introduction}
Prediction markets aggregate beliefs through financially consequential choices. Participants trade contracts tied to future events, and prices summarize the terms on which they are willing to transact. The attraction is that dispersed information can enter a common forecast without requiring any individual to possess the full picture \citep{wolfers2004prediction}. On Polymarket, trading takes place alongside public comment threads. The two records reveal different aspects of collective behavior: trades allocate money, while comments reveal attention, emotional tone, and expressed support for an outcome.

We ask whether discussion helps predict what a market will do next. More comments could precede more trading without indicating which outcome attracts the buying. Expressed support could anticipate buying direction without forecasting a change in the leading outcome. An active thread can also accompany a price that remains near certainty. We therefore distinguish trading activity, buying direction, and changes in leadership, evaluating each against information already present in the trading record.

Polymarket supports trading in outcome tokens. In a binary market, the token corresponding to the realized outcome can be redeemed for one dollar; the other becomes worthless. Prices are commonly interpreted as probabilities, subject to the information and incentives underlying trading \citep{wolfers2004prediction,polymarketMarkets}. An off-chain central limit order book matches orders, while smart contracts settle collateral and outcome-token transfers on the Polygon blockchain \citep{polymarketExchange,polymarketTrading}. Makers supply resting orders and takers execute against them; either role can buy or sell. On fee-enabled markets, taker fees provide platform revenue and partly fund maker rebates, with the documented schedule varying by category and price \citep{polymarketFees}. The market universe includes cryptocurrency, sports, finance, economics, politics, culture, technology, and weather.

We focus on non-political markets to examine collective forecasting of external outcomes rather than outcomes determined by participants' votes. This complements work on the wisdom of crowds in political prediction \citep{sethi2025political}. We retain single-market events so that each discussion thread maps to one price series, exclude recurring series and autogenerated fixtures, and require sufficient discussion to construct temporal features. Contestedness distinguishes sustained uncertainty or repricing from active discussion around largely certain outcomes.

The analysis has two stages. An exploratory sweep compares comment and market signals across temporal resolutions, leads, comment thresholds, and price bands. Six experiments then evaluate the selected settings through trading-only, comment-only, and combined models. Whole-market holdouts assess transfer across markets, while baselines and operating-point comparisons distinguish useful rankings from reliable decisions at a common threshold.

Three findings organize the paper. Attention nearly matches trading history for heavy-activity prediction and has its greatest relative advantage in short markets. Comment content supplies buying-direction information, with modest gains when combined with flow history. Leadership changes depend primarily on market state: discussion shifts against leaders before reversals, but this aggregate pattern does not yield a clear additional forecasting benefit. We interpret these as predictive associations. Shared developments may affect both discussion and trading; our observational design does not identify the direction or presence of causal effects between them.

%% file: related_work.tex
\section{Related Work}
\label{sec:related-work}
\paragraph{Collective judgment and market discussion.}
Prediction markets can aggregate dispersed information, but their accuracy depends on market design and participation \citep{wolfers2004prediction}. Favorite--longshot bias illustrates departures from probability assessment \citep{thaler1988parimutuel}. \citet{sethi2025political} compare statistical forecasts with Polymarket through a profitability-based evaluation, showing that relative performance depends on the contract. \citet{gomez2026accuracy} attribute forecasting accuracy to a persistently skilled minority, while \citet{akey2026wins} examine the distribution of trading outcomes. These studies concern information aggregation through trades; we examine what public discussion predicts about subsequent market behavior.

\citet{dahlke2026electoral} jointly describe trading, commenting, and reacting, finding concentrated and partly overlapping participation and examining temporal ordering within wallet--event pairs. Our unit is instead the market-time block, and our question is whether aggregated discussion adds to trading-history forecasts. \citet{mbrice2026stance} studies stance detection in trader commentary; we use stance as an input to prediction. \citet{kim2026risk} combine lead--lag screening with language-model filtering of relationships between event descriptions. Their signals link market probability series; ours pair discussion with later trading and price outcomes.

\paragraph{Text and collective attention.}
\citet{tetlock2007content} links media pessimism to market movements, while \citet{mao2011predicting} compare survey, news, social-media, and search signals. \citet{keskin2020information} examine information transfer between sentiment and cryptocurrency prices. These precedents motivate temporal analysis without equating prediction with causation. Collective-attention research supplies a complementary perspective: \citet{hogg2012digg} model popularity through visibility and user interest; \citet{burghardt2017myopia} show how cognitive load shapes collective evaluation; \citet{he2021attention} study shifts of attention between topics; and \citet{he2022changepoint} use changepoints to identify candidate natural experiments. Together, these works motivate separating participation dynamics from the beliefs expressed in a discussion.

\paragraph{Signal screening and temporal dependence.}
Our sweep faces the search problem familiar from finance's factor zoo \citep{harvey2016cross,jensen2023replication}: searching many plausible signals complicates interpretation of apparent predictability. Data-snooping tests and false-discovery control address this problem under their respective assumptions \citep{white2000reality,benjamini1995fdr}. Genetic association research similarly separates discovery from independent replication \citep{nci2007replicating}, while specification-curve analysis makes variation across analytical choices visible \citep{simonsohn2020specification}. We use screening to choose experimental settings, without claiming multiplicity-calibrated discoveries or an independent replication cohort. Serial dependence also affects correlation inference \citep{pyper1998autocorrelation}; held-out-market prediction addresses transfer but does not retrospectively calibrate the sweep's tests.

%% file: data_selection.tex
\section{Data Selection}
\label{sec:data-selection}
We focused on non-political Polymarket markets with active discussion and uncertain outcomes. Because threads belong to events, we retained only single-market events, giving each thread one corresponding price series. We excluded recurring series, whose counts could be inherited from shared threads, and automatically generated fixture markets. Candidates required more than 100 platform-reported comments.

We characterized contestedness through sustained uncertainty and repricing. Each market's lifetime was normalized to 200 bins. A market was contested if its winning-side price had binary entropy of at least 0.5 in at least 20\% of bins, or total log-odds variance of at least 40. Four uncontested markets were retained in the collected corpus to permit comparison with active discussions around largely certain outcomes. The numerical criterion described contestedness; it did not exclude markets from the reported analyses.

The data have four files per market: a complete trade history and sentiment, toxicity, and stance annotation files. The three annotation sets match by comment identifier and cover the same 63,720 retrieved and scored comments in the collected corpus. The sweep used 79 markets. Temporal truncation and feature warm-up left 18,399 six-hour blocks across 76 markets; Experiment~6 used 18,323 labelled blocks across 74 markets because two panel markets lacked stance labels for that run. Section~\ref{sec:methods} describes these analysis-specific requirements.

%% file: sweep.tex
\section{Correlation Sweep}
\label{sec:sweep}
Before classification, we screened associations to select temporal resolution, comment thresholds, leads, and feature families. The sweep covered 79 markets with complete trade histories and comment measures. Trades and comments were aligned to a fixed UTC clock. A winning token required a final trade price of at least \$0.95; an extreme-longshot market failed this criterion and could not be processed. Histories were truncated when the winning token's daily close first reached \$0.99.

We crossed 15 comment measures with 13 market measures, bin durations of 6, 12, and 24 hours, minimum comment counts of 1, 5, and 10, and leads of one, two, and three bins. Seven price settings comprised no restriction and the bands 0--25, 25--45, 45--55, 55--75, 75--90, and 90--100 cents. Price restrictions used the comment bin's winning-token price. Leads were always forward: a one-bin lead paired comments with market activity in the following bin.

Comment families comprised attention (counts, distinct commenters, reactions), hostility (toxicity and harassment scores), and tone (sentiment levels, dispersion, intensity, and polarity shares). Market families comprised activity (trade count, volume, trade size, and timing), price (token prices and log-odds), and direction (buy-flow and aggressor imbalance). A \emph{cell} was one complete configuration for one market; a \emph{recipe} was the corresponding configuration across markets. Spearman ranks reduced the influence of unusually large activity bins. A cell cleared the screen if it had at least 15 observations and analytic $p<0.01$.

The grid contained 34,938,540 cells; 2,388,347 were computable. Whole-history configurations without phase or liveness filters yielded 246,739 eligible cells across 57 markets. The other 22 markets lacked 15 bins containing comments. Attention cleared most often against activity, followed by hostility and tone (Table~\ref{tab:sweep}). At the six-hour, one-comment, one-bin-lead setting without price restriction, distinct-commenter counts predicted subsequent trade count in 32 of 57 markets; all clearing correlations were positive, with median $r=0.58$. Insult scores cleared in 15 markets, all positive, with median $r=0.41$.

\begin{table}[t]
\centering
\begin{tabular}{lrrr}
\hline
Comment family & Activity & Price & Direction \\
\hline
Attention & 33.0 & 14.7 & 7.2 \\
Hostility & 8.8 & 5.0 & 2.7 \\
Tone & 2.6 & 2.1 & 1.8 \\
\hline
\end{tabular}
\caption{Cells clearing the correlation screen (\%) by feature family. Rates describe eligible configurations, not predictive accuracy or calibrated discovery rates.}
\label{tab:sweep}
\end{table}

Price and directional associations were less uniform. Attention--price cells had almost equal positive and negative signs. Of 415 clearing hostility--direction cells, 257 occurred in live stretches (58\% positive), 66 in settled stretches (15\% positive), and 92 across both states (42\% positive). Mean sentiment against buy-flow imbalance cleared in none of the 57 testable markets at the representative setting. Among 315 recipes clearing in at least ten markets, 173 had identical signs across clearing markets, all involving attention or hostility against activity.

Short leads concentrated the strongest screening rates. Requiring more comments did not improve yield, and price restrictions reduced the eligible sample. Six-hour bins supplied 7,775 clearing cells, compared with 6,334 at 12 hours and 4,503 at 24 hours. Activity associations were more common in live than settled stretches for attention (33\% versus 24.2\%); phase of market life provided little separation. These findings motivated six-hour bins, at least one comment where discussion was required, short forward targets, and no general price-band restriction. Liveness was tested for burst prediction, while hostility and tone remained available to classifiers despite weaker screening results. Stance was annotated after the sweep.

Adjacent bins and overlapping configurations are dependent. No permutation null was computed, so neither the nominal $p$-values nor the clearing frequencies estimate a calibrated number of discoveries. The sweep selects settings within this corpus; the following experiments assess prediction under those settings, rather than providing an independent replication of the entire discovery process.

%% file: methods.tex
\section{Experimental Design}
\label{sec:methods}
\subsection{Temporal Construction and Features}
The six-hour panel used blocks beginning at 00:00, 06:00, 12:00, and 18:00 UTC. Following the resolution cutoff in Section~\ref{sec:sweep}, we removed the first four blocks for rolling-feature warm-up. Three sweep markets then had no usable blocks, leaving 76 markets. Features were computed on each complete block sequence before experiment-specific filtering, so trailing histories included preceding uncommented blocks. Features describe the current block and its available history; targets describe subsequent blocks.

A block was \emph{live} if, over the preceding week, price crossed \$0.50 or spanned at least \$0.10. This measures recent movement, whereas contestedness describes uncertainty. Markets lasting at most 21 days before cutoff were designated short. Directional-flow targets and stance were oriented toward the first-listed outcome. Leader-oriented features instead express support for whichever outcome currently leads.

\paragraph{Comment features.}
The activity experiments used five attention features: comment presence, log comment count, log distinct-author count, log trailing 24-hour comment count, and current count divided by its trailing 48-hour mean. Sentiment comprised a RoBERTa compound score and positive, negative, and neutral shares. Social-media language models and benchmarks provide the methodological background \citep{barbieri2020tweeteval,loureiro2022timelms}. Toxicity used Detoxify threat, insult, obscenity, and identity-attack scores \citep{hanu2020detoxify}, their four standardized counterparts, and overall hostility. Standardization used each market's strictly preceding expanding history, with at least six periods; missing standardized scores were filled with zero. Overall hostility was the maximum of the four standardized scores. Missing raw toxicity values were zero; undefined count and trade ratios were filled with one.

Claude Sonnet 5 generated bullish, bearish, or neutral stance labels and confidence scores using each market's verbatim resolution criteria. Market-specific prompts were frozen across batches. Seven block-level features comprised the three label shares, net stance (bullish minus bearish share), confidence-weighted net stance, log labelled-comment count, and trailing four-block mean net stance. Confidence weighting multiplies confidence by $+1$, $-1$, or zero for bullish, bearish, or neutral labels. Stance follows the first-listed outcome independently of eventual resolution.

\paragraph{Trading and market-state features.}
Burst models used log trade count, its trailing 48-hour surge ratio, and a bursting-now indicator based on the expanding 80th percentile of preceding counts. The activity-change model added trailing 24-hour mean absolute price movement. Flow models used current buy-flow imbalance, its four-block average, and log trade count. Buy-flow imbalance contrasts dollars buying the two outcomes, divided by their sum. Tick-based aggressor imbalance provides a distinct directional measure, following the microstructure distinction between trade direction and order imbalance \citep{lee1991direction,chordia2002imbalance}.

The leadership experiment used 36 features: 19 comment and 17 market-state signals. Its comment set comprised six attention features (raw and log counts of comments and authors, log trailing comment count, and the count ratio), four sentiment features, and nine toxicity features. Market state comprised raw and log trade count, trade-count ratio, dollar volume, bursting-now and large-movement indicators, trailing log-odds movement, distance from \$0.50, $p(1-p)$, liveness, and seven leader-oriented signals: current and trailing buy-flow imbalance, current and trailing aggressor imbalance, and current movement sign, signed magnitude, and trailing signed magnitude. Burst and large-movement indicators used preceding 80th and 85th percentiles with eight historical blocks. Leader orientation multiplied reference-token features by $+1$ or $-1$ according to its current side of \$0.50. Stance was tested separately rather than included among the 36 features.

\subsection{Tasks and Evaluation}
Experiments~1--3 predict heavy trading. Experiment~1 asks whether any of the next three blocks exceeds the market's whole-life trade-count 80th percentile. This defines the target; the feature-side bursting indicator uses preceding history only. Applying live and comment filters to 18,399 blocks leaves 7,133 and then 2,462 blocks across 65 markets, of which 51 satisfy scoring requirements. Experiment~2 groups these predictions by market duration. Experiment~3 predicts a burst in the next block without a live filter, using 3,801 commented blocks across 74 markets, with 62 scored.

Experiment~4 predicts the sign of mean buy-flow imbalance in the next three blocks. It uses 3,699 commented blocks with defined direction across 74 markets, with 53 scored. Experiment~5 predicts whether mean trade count in the next three blocks exceeds the current count, using 3,691 commented blocks across 74 markets. Experiment~6 predicts a change of leader within the next four blocks using 18,323 labelled blocks across 74 markets. Two panel markets lacked stance labels for that run. The target prevalence is 0.078; no comment filter is imposed.

Experiments~1--4 use L2-regularized logistic regression with $C=0.5$ and standardization fitted on training markets. Experiment~1 also uses 300 shallow boosted trees. Experiment~5 uses 300 boosted trees of depth three, learning rate 0.05, and 80\% row and column sampling; Experiment~6 uses boosted trees and feature ablations. The primary evaluation in Experiments~1--4 holds out one entire market at a time. Scoring requires at least ten qualifying blocks and both target classes; markets not satisfying these conditions can still contribute training observations. Unless stated otherwise, scores are averaged across scored markets.

Experiment~4 additionally uses 20 random splits into 58 training and 16 test markets. Experiment~5 compares 20 splits into 56 training and 18 test markets with 20 splits into 2,953 training and 738 test blocks. The latter measures prediction on additional observations from markets represented in training. Experiment~6 uses five folds grouped by market. Its horizon ablations use a fixed seed-0 holdout of 20 of the 76 panel markets: 2,391 blocks with a defined four-block label, including 668 commented blocks. Reported $\pm$ values are population standard deviations across repetitions or folds.

We evaluate PR-AUC against mean prevalence and ROC-AUC against 0.5. Experiment~3 also evaluates random scores over 200 seeds. Decision comparisons use training-majority accuracy, held-out-market majority accuracy as a hindsight benchmark, and a current-burst persistence rule. Precision, recall, and lift are reported at probability above 0.5 and in the highest-scoring decile. Pooled operating-point prevalence can differ from mean market prevalence; repeated-split counts include repeated test predictions. Whole-market holdouts separate markets within each fit, while block-level splits answer a different transfer question.

%% file: results.tex
\section{Results}
\label{sec:results}
\subsection{Experiments 1--3: Attention and Trading Activity}
\paragraph{Heavy trading on live, commented blocks.}
Attention nearly matched trading history in Experiment~1: PR-AUC was 0.786 versus 0.790, against mean prevalence 0.606; their combination achieved 0.793 (Table~\ref{tab:activity}). Attention exceeded trading in 23 of 51 scored markets. Under boosting, the combination improved from trading's 0.767 to 0.781 and exceeded trading in 29 markets. Toxicity and sentiment alone were weaker, at 0.663 and 0.626 with logistic regression; adding toxicity did not improve on trading alone.

Ranking performance did not imply equally strong decisions at probability 0.5. Attention accuracy was 0.612, compared with 0.685 for trading and 0.684 for their combination, against training-majority accuracy 0.606. On 2,393 scored blocks, attention flagged 1,971 at that threshold and achieved recall 0.875 but precision 0.660. Restricting predictions to the highest-scoring decile increased attention precision to 0.745; the combination reached 0.833, correctly identifying 199 of 239 flags at recall 0.134. Attention helped order blocks, while the combined forecast concentrated the strongest burst predictions.

Removing liveness filtering reduced attention's relative strength: approximately 70\% of all blocks lacked comments, and trading outperformed attention by 0.745 versus 0.685 without the comment filter. The parity also disappeared at daily resolution (0.775 attention versus 0.811 trading). Raising comment thresholds improved attention by only 0.012 at five comments and 0.045 at ten, while changing which blocks and markets could be scored. Much of the comment signal was already present when a single comment occurred.

\begin{table}[t]
\centering
\small
\begin{tabular}{lrrrr}
\hline
Setting & Chance & Attention & Trading & Both \\
\hline
E1: live, 18h & .606 & .786 & .790 & \textbf{.793} \\
E3: all, 6h & .387 & .637 & \textbf{.692} & .673 \\
E5: markets & .484 & .530 & .734 & .734 \\
E5: blocks & .483 & .534 & .745 & .750 \\
\hline
\end{tabular}
\caption{PR-AUC for activity targets. All rows require comments. E1 and E3 predict bursts; E5 predicts increases relative to current activity. E5 ``Both'' combines trading and attention. E3's empirical random-score reference is .447. Settings differ in target, sample, and evaluation.}
\label{tab:activity}
\end{table}

\paragraph{Market duration and the attention advantage.}
Experiment~2 found that markets favoring attention were shorter and more densely discussed. Median duration was 15.0 versus 76.5 days for markets favoring trading, and median comments per day were 15.1 versus 4.2. Mann--Whitney $p$-values were below 0.001 and 0.020, respectively. Total comments and trades did not distinguish the groups. The advantage concerned the timing and concentration of discussion rather than accumulated market size.

\begin{figure}[t]
\centering
\includegraphics[width=\columnwidth]{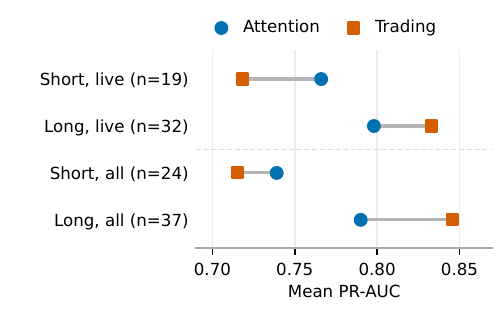}
\caption{Attention's relative advantage is concentrated in short markets (at most 21 days). Points are reported mean PR-AUC values from Experiment~2; $n$ denotes scored markets. All groups require comments. The ordering persists without the live filter.}
\label{fig:duration}
\end{figure}

Figure~\ref{fig:duration} shows the duration comparison using the same leave-one-market-out predictions, grouped after prediction. On 19 live short markets, attention achieved PR-AUC 0.766 versus trading's 0.718. On 32 live long markets, trading was stronger, 0.833 versus 0.798. Without the live filter, attention retained its advantage on 24 short markets (0.739 versus 0.715), while trading led on 37 long markets (0.846 versus 0.790). Attention remained informative in long markets; its relative advantage narrowed as the trading baseline strengthened.

At the top decile, combined features achieved precision 0.825 on short markets (33 of 40 flags), compared with 0.775 for trading and 0.625 for attention. On long markets, trading alone reached 0.975 (195 of 200 flags). Attention's short-market recall at probability 0.5 was 0.990, but its precision of 0.524 was close to pooled prevalence 0.520. Thus, its ranking advantage did not translate into a well-calibrated common decision threshold.

\paragraph{Predicting the next six hours.}
Experiment~3 removed the live filter and shortened the target to the next block. Attention achieved PR-AUC 0.637, trading 0.692, and their combination 0.673. They exceeded each market's prevalence in 59, 62, and 61 of the 62 scored markets. Random scores averaged $0.447\pm0.012$ across 200 seeds, with minimum 0.407, above mean prevalence 0.387. This finite-sample effect makes 0.447 the relevant empirical random-ranking reference; all three models exceeded it.

Trading also exceeded decision baselines: accuracy 0.710 versus 0.613 for training majority and 0.689 for held-out-market majority. Attention accuracy was 0.594. On 3,741 scored blocks, trading's top decile correctly identified 260 of 374 flags (precision 0.695), compared with attention precision 0.529. The attention signal persisted across the broader sample and shorter horizon, while trading supplied the stronger next-block forecast.

\subsection{Experiment 4: Direction of Buying}
All four comment families ranked future buying direction above ROC chance on average (Table~\ref{tab:direction}). Toxicity was strongest alone, with ROC-AUC 0.586 and above-chance ranking in 43 of 53 markets. Stance achieved 0.580, attention 0.571, and sentiment 0.540. These results show directional structure that was weak or inconsistent in the individual-market sweep.

The strongest combined representation added attention, sentiment, and stance to flow history, increasing ROC-AUC from 0.780 to 0.788 and exceeding chance in 51 of 53 markets. Accuracy rose from 0.805 to 0.812. Toxicity did not improve this combination, although it was the strongest comment family alone. At the top decile, the strongest combined model correctly identified 329 of 341 flags (precision 0.965, recall 0.207), compared with flow-only precision 0.962. At probability 0.5, its precision was 0.816 and recall 0.824. Stance alone ranked outcomes above chance but achieved accuracy 0.546 and exceeded the held-out-majority baseline in only 7 markets.

\begin{table}[t]
\centering
\small
\begin{tabular}{lrrr}
\hline
Features & ROC-AUC & Above .5 & Accuracy \\
\hline
Toxicity & .586 & 43/53 & .496 \\
Stance & .580 & 34/53 & .546 \\
Attention & .571 & 34/53 & .554 \\
Sentiment & .540 & 33/53 & .496 \\
Flow history & .780 & 49/53 & .805 \\
Flow + stance & .779 & 51/53 & .803 \\
Flow + A, S, stance & \textbf{.788} & 51/53 & \textbf{.812} \\
\hline
\end{tabular}
\caption{Experiment~4: mean held-out-market buying-direction performance. A and S denote attention and sentiment. ``Above .5'' counts markets with ROC-AUC above chance.}
\label{tab:direction}
\end{table}

Across 20 repeated market splits, stance achieved ROC-AUC $0.600\pm0.065$; flow history achieved $0.908\pm0.024$, and adding stance yielded $0.909\pm0.033$. Restricting to live blocks reduced stance performance to 0.584 across 44 markets. Daily aggregation reduced it to 0.547 across 34 scored markets, compared with 0.805 for flow and 0.802 for flow with stance. Directional comment information therefore depended on temporal resolution.

Collapsing each of 74 markets to a single observation removed the predictive pattern. Whole-life net stance correlated with whole-life flow direction at $\rho=0.15$ ($p=0.202$), and market-level stance classifiers achieved ROC-AUC 0.512 for flow direction and 0.485 for the winning outcome. The useful signal concerned timing within a market, rather than which markets had more optimistic discussions overall.

\subsection{Experiment 5: Increases in Activity}
Predicting increases relative to the current block improved agreement between rankings and thresholded decisions. On unseen markets, trading achieved PR-AUC $0.734\pm0.027$, ROC-AUC $0.737\pm0.024$, and accuracy $0.677\pm0.019$, compared with majority accuracy 0.516. Attention was the strongest comment family but remained weaker: PR-AUC $0.530\pm0.024$. Adding attention left PR-AUC at 0.734 and increased ROC-AUC to 0.740. Sentiment, stance, and their combinations offered little consistent improvement.

Under random block splits, trading achieved PR-AUC $0.745\pm0.014$, increasing to $0.750\pm0.015$ with attention. Pooled market-split trading predictions reached top-decile precision 0.871 (1,575 of 1,808 flags), while trading with attention reached 0.865. Under block splits, corresponding precisions were 0.885 (1,306 of 1,476) and 0.888 (1,311 of 1,476). These represent lifts of approximately 1.8 over prevalence; attention alone yielded approximately 1.2.

Paired-block constructions tested overlap between samples and target windows. Excluding starting blocks previously used in targets reduced the sample from 3,637 to 1,822. Trading PR-AUC changed from 0.652 to 0.661; trading with stance achieved 0.671 under the stricter construction. The ordering remained dominated by trading history. Adding liveness and contestedness as features changed ROC-AUC by less than 0.001, providing little beyond the existing movement and activity features.

\subsection{Experiment 6: Changes in Market Leadership}
Across five market-grouped folds, the full model achieved mean PR-AUC 0.420 and pooled PR-AUC 0.403, against pooled prevalence 0.078; pooled ROC-AUC was 0.868. ROC-AUC exceeded each fold's chance reference in all five folds. Predicting no change would achieve 92.2\% accuracy, making rare-event ranking and operating points more informative here.

Feature ablations showed that market state supplied most predictive value (Table~\ref{tab:leadership}). On the fixed holdout, four-block PR-AUC was 0.517 for the full model, 0.515 for market state, 0.511 for price distance alone, and 0.158 for comments. On commented blocks, corresponding values were 0.386, 0.407, 0.428, and 0.233. Split-based feature importance similarly emphasized distance from even odds, $p(1-p)$, liveness, and recent movement; these describe proximity to a crossing and current market dynamics.

\begin{table}[t]
\centering
\small
\begin{tabular}{lrrr}
\hline
Features & All, 6h & All, 24h & Comments, 24h \\
\hline
Full & .242 & .517 & .386 \\
Market state & .260 & .515 & .407 \\
Comment signals & .057 & .158 & .233 \\
Price distance & .259 & .511 & .428 \\
\hline
Prevalence & .042 & .120 & .162 \\
Blocks & 2,391 & 2,391 & 668 \\
\hline
\end{tabular}
\caption{Experiment~6: PR-AUC on the fixed 20-market holdout. Both horizons use blocks with defined four-block labels. ``Comments'' restricts to commented blocks. Stance is excluded and tested separately.}
\label{tab:leadership}
\end{table}

On all 18,323 blocks, the market-state model flagged 745 at probability 0.5 and captured 408 reversals: precision 0.548, recall 0.287, and lift 7.1. Its top decile captured 736 reversals among 1,832 flags, giving precision 0.402, recall 0.518, and lift 5.2. Combined features produced top-decile precision 0.398; comments alone produced 0.092. Market state identified approximately half the reversals while flagging one tenth of observations.

\paragraph{Stance before a reversal.}
On 3,801 commented blocks, mean stance toward the current leader was $-0.080$ within a day before reversal and $+0.025$ otherwise. This contrast was not explained only by proximity to \$0.50: among 27 markets with sufficient observations more than \$0.20 from even odds in both groups, 23 showed more negative stance before reversals (reported two-sided exact binomial $p\approx0.0003$). Typical within-market gap was 0.235, larger than the pooled gap. Opposition to the leader preceded the change rather than simply accompanying it.

Individual-block stance nevertheless remained variable. In grouped evaluation on commented blocks, stance alone achieved PR-AUC $0.121\pm0.043$ against prevalence 0.101 and ROC-AUC 0.486. Market state achieved PR-AUC $0.394\pm0.109$; adding stance yielded $0.414\pm0.107$, with ROC-AUC changing from 0.852 to 0.853. The PR-AUC increase was smaller than fold variation, so improvement was not established with confidence. At the top decile, adding stance captured 178 rather than 172 reversals among 380 flags, increasing precision from 0.453 to 0.468. The aggregate anticipatory pattern did not produce a clear improvement in forecasting individual reversals.

%% file: conclusion.tex
\section{Discussion and Conclusion}
\label{sec:conclusion}
Public discussion carries different information about different aspects of a prediction market. Attention is informative about subsequent participation: it nearly matches trading history for heavy activity on live, commented blocks, and its relative advantage is concentrated in short markets. This advantage concerns when discussion and trading become active, rather than the total size of a market's audience. It persists without liveness filtering, while trading history becomes the stronger baseline in longer markets.

Comment content has a different role. Toxicity and stance rank subsequent buying direction above chance, and combining selected comment features with flow history produces modest improvements. These patterns weaken under daily aggregation and disappear when each market is reduced to a single stance summary. Their predictive value is temporal and local to the market, rather than a general relationship between optimistic discussion and favorable outcomes.

Leadership changes draw most of their predictability from market state. Discussion becomes less supportive of a leader before it loses that position, including away from even odds, yet stance varies too much across individual blocks to deliver a clear additional forecast improvement. An aggregate behavioral pattern and an effective event-level predictor are different findings; the results support the former more strongly for stance before reversals.

The scope of these conclusions is the selected, discussion-rich, non-political corpus and the reported experimental settings. The exploratory sweep uses dependent observations and no calibrated permutation null; the evaluated markets also informed setting selection. Whole-market holdouts assess transfer within that design rather than independent discovery replication. Language annotations are model-derived, and neither the source-domain benchmarks nor the forecasting comparisons establish their accuracy for every comment. The study does not identify causal effects, private information, or trading profitability.

Together, the experiments show where discussion complements the trading record and where it contributes little beyond existing market dynamics. Attention helps anticipate participation, content helps distinguish buying direction, and prices remain central to forecasting a change of leader. These distinctions make discussion a useful additional observation of collective forecasting while specifying the limits of what it predicts.

%% file: acknowledgements.tex
\section*{Acknowledgements}
\paragraph{Generative AI disclosure.}
Claude (Anthropic) assisted with analysis code, experimental reports, and drafting, and generated comment-level stance labels and confidence scores from market-specific resolution criteria. ChatGPT (OpenAI) assisted with manuscript drafting and revision, LaTeX text and tables, the results figure, reference checking, and BibTeX preparation \citep{anthropicClaude,openaiChatGPT}. The author retains responsibility for research decisions, interpretation, accuracy, and the final manuscript.